\documentclass[9pt,twocolumn,twoside]{opticajnl}
\journal{opticajournal}
\setboolean{shortarticle}{true}

\title{Theory of Airy-Beam Wavefront Sensing}

\author[1,*]{Surya Kamal}
\affil[1]{Chester F. Carlson Center for Imaging Science, Rochester Institute of Technology, Rochester, NY 14623, USA}
\affil[*]{Corresponding author: skcis@rit.edu}

\begin{abstract}
A new wavefront sensing framework based on the interference of counter-accelerating Airy beams is presented. An engineered cubic phase function transforms a single Gaussian beam into two coherent Airy beams with opposite transverse accelerations, which self-recombine during free-space propagation to form an interference pattern encoding the aberrated incident wavefront. Numerical simulations demonstrate accurate recovery of the aberrated exit-pupil phase across the aberration sets considered, with a root-mean-square phase reconstruction error below $10^{-6} \mathrm{\;rad}$ for the aberration set shown and consistently low errors for the other sets, while maintaining accurate recovery in the presence of measurement noise. The approach eliminates collection optics and enables a compact wavefront-sensing architecture for applications where conventional interferometric systems are difficult to implement, including electron-optical systems, augmented and virtual reality systems.
\end{abstract}

\setboolean{displaycopyright}{false}

\begin{document}
\maketitle

Wavefront sensing is a fundamental capability in optical imaging and has found widespread applications in microscopy, astronomy, laser systems, and adaptive optics for diffraction-limited imaging \cite{hampson2021adaptive}. Existing wavefront-sensing approaches generally rely on direct wavefront sensors, such as the Shack--Hartmann wavefront sensor (SHWS), non-interferometric phase retrieval from intensity measurements, or interferometric techniques including digital holography \cite{hampson2021adaptive,mugnier2006phase,tahara2018digital}. SHWSs measure local wavefront slopes across an array of subapertures, whereas intensity-based phase-retrieval methods generally require multiple measurements or diversity constraints in the image or pupil plane \cite{hampson2021adaptive,mugnier2006phase}. Many interferometric techniques provide high phase sensitivity but require beam splitting and subsequent recombination to generate an interference pattern \cite{tahara2018digital}. In compact optical architectures, as well as matter-wave optical systems (electrons, ions, etc.), these additional optical elements can increase system complexity, introduce additional aberrations, or become impractical to implement. Consequently, there is a need for wavefront-sensing methods that eliminate the requirement for collection optics while preserving the phase sensitivity of interferometric measurements.

Airy beams offer a particularly promising route toward collection-optics-free interferometry because their characteristic curved trajectories can enable spatially separated beams to naturally overlap during propagation. Since their theoretical prediction \cite{balazs1979nonspreading} and experimental realization \cite{siviloglou2007observation}, Airy beams have attracted extensive interest in a broad range of wave-based systems, including photons \cite{siviloglou2007observation}, electrons \cite{voloch2013generation}, and neutrons \cite{sarenac2025generation}. In addition to their diffraction-resistant propagation, self-healing, and transverse acceleration, Airy beams have been explored for applications in optical manipulation, microscopy, imaging, communications, and particle-beam engineering \cite{efremidis2019airy}. The interference of Airy beams has also been investigated, including the near-field interference of spatially separated Airy modes arising from their transverse self-acceleration \cite{wang2025interference}. In particular, counter-accelerating Airy beams can follow opposing curved trajectories and subsequently overlap, providing a means for self-recombination and interference without conventional collection and beam-recombination optics. These properties make counter-accelerating Airy beams an attractive platform for interferometric wavefront sensing, particularly in compact optical and matter-wave systems where conventional interferometric architectures can be challenging to implement.

In this Letter, we present a wavefront-sensing framework based on the interference of counter-accelerating Airy beams. An engineered cubic phase function is used to transform a Gaussian beam into two coherent Airy beams with opposite transverse accelerations, whose curved propagation trajectories enable self-recombination without collection optics. The resulting interference pattern encodes the aberrated exit-pupil wavefront, which is recovered using a four-step interferometric phase-retrieval procedure followed by model-based reconstruction of the aberration coefficients. Wave-optical simulations demonstrate accurate recovery of the aberrated wavefront, with low root-mean-square phase reconstruction error and robust performance in the presence of measurement noise.


\begin{figure}[ht]
\centering
\includegraphics[width=\linewidth]{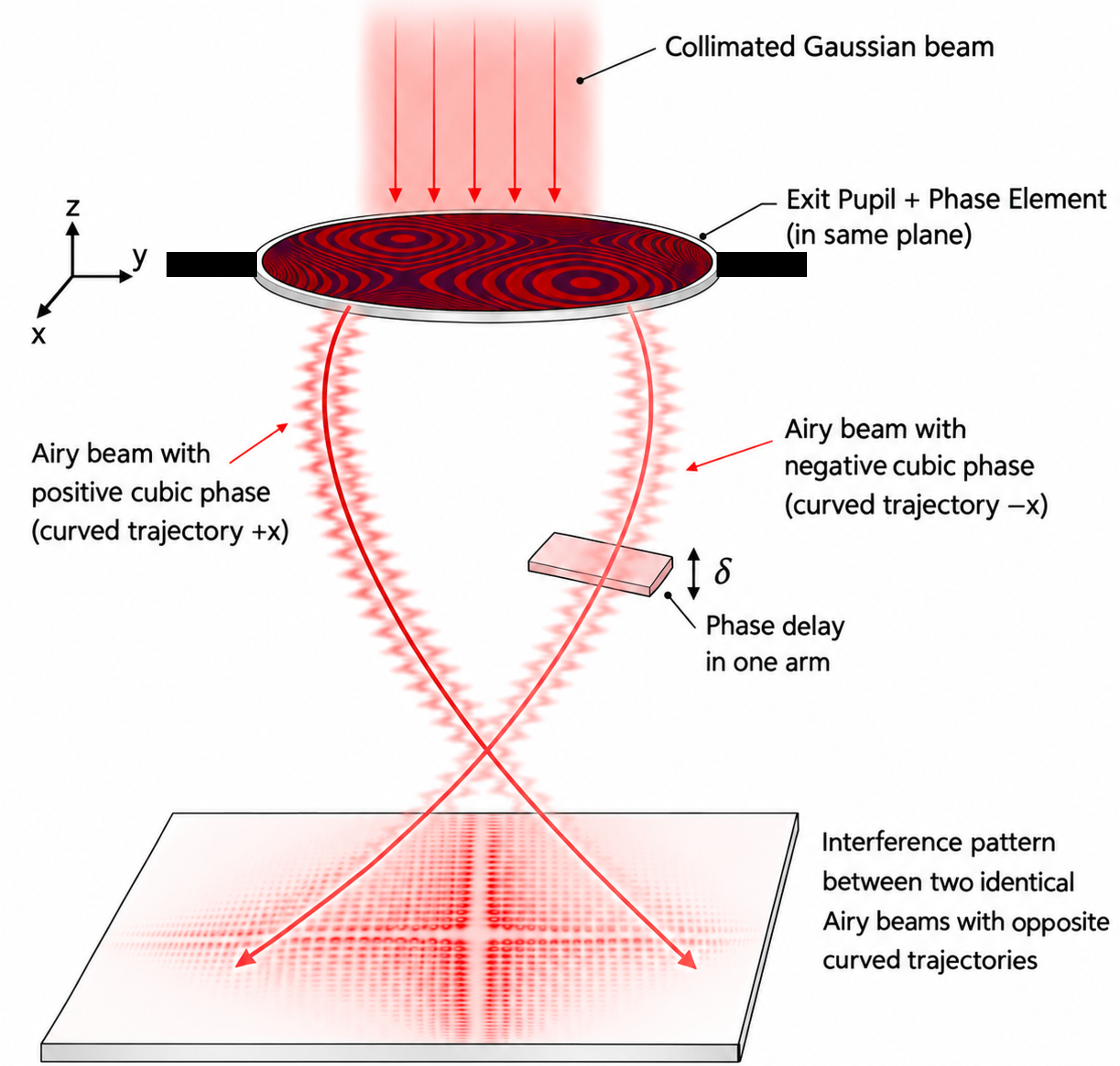}
\caption{
Conceptual architecture of the proposed Airy-beam wavefront sensor. A Gaussian beam illuminates an aberrated exit pupil containing the optical phase structure of the system. An engineered cubic-phase element generates two coherent channels with opposite cubic phase. The corresponding Airy fields propagate to the Fourier plane, where they interfere and the resulting intensity is recorded for phase recovery.
}
\label{fig:schematic}
\end{figure}

\textit{Cubic-phase encoding:} Figure~\ref{fig:schematic} illustrates the proposed wavefront-sensing architecture. A Gaussian beam illuminates an exit pupil (EP) containing the optical phase of the system under test. The pupil may include the phase contribution of the imaging optics together with an unknown aberration,
\begin{equation}
E_{\mathrm{p}}(x,y)
=
A(x,y)E_{\mathrm{G}}(x,y)
\exp\left[i\Phi_{\mathrm{p}}(x,y)\right],
\label{eq:pupil_field}
\end{equation}
where $A(x,y)$ denotes the pupil amplitude, $E_{\mathrm{G}}(x,y)$ is the incident Gaussian field, and $\Phi_{\mathrm{p}}(x,y)$ contains the phase introduced by the optical system and the unknown aberration.

The phase to be recovered is written as
\begin{equation}
\Phi_{\mathrm{p}}(x,y)
=
\Phi_{0}(x,y)
+
\Phi_{\mathrm{ab}}(x,y),
\label{eq:pupil_phase}
\end{equation}
where $\Phi_{0}$ represents the known or calibrated nominal phase and $\Phi_{\mathrm{ab}}$ is the unknown aberration.

An engineered phase element introduces two oppositely signed cubic phase functions,
\begin{equation}
\Phi_{c,\pm}(x,y)
=
\pm
\alpha
\left[
\left(\frac{x}{R}\right)^3
+
\left(\frac{y}{R}\right)^3
\right],
\label{eq:cubic_phase}
\end{equation}
where $\alpha$ controls the cubic-phase strength and $R$ is the pupil radius. Cubic phase masks are a standard means of generating and controlling Airy-type optical fields \cite{siviloglou2007accelerating}.

The two coherent fields immediately following the engineered phase element are therefore
\begin{equation}
E_{\pm}(x,y)
=
E_{\mathrm{p}}(x,y)
\exp\left[
\pm i\Phi_c(x,y)
\right].
\label{eq:airy_channels}
\end{equation}

The important feature of Eq.~\eqref{eq:airy_channels} is that the two channels originate from the same aberrated pupil. Thus, the unknown aberration is common to both coherent fields rather than being assigned to only one interference arm.

The opposite cubic phases produce conjugate Airy-type propagation responses. The resulting fields at the measurement plane are
\begin{equation}
U_{\pm}(x,y)
=
\mathcal{P}_{z}
\left\{
E_{\mathrm{p}}(x,y)
 e^{\pm i\Phi_c(x,y)}
\right\},
\label{eq:propagated_fields}
\end{equation}
where $\mathcal{P}_{z}$ denotes the propagation operator from the exit pupil to the Fourier-plane measurement plane.

For numerical propagation, the angular-spectrum operator is used,
\begin{equation}
U_{\pm}(x,y,z)
=
\mathcal{F}^{-1}
\left\{
\mathcal{F}
\left[
E_{\mathrm{p}}e^{\pm i\Phi_c}
\right]
H(k_x,k_y)
\right\},
\label{eq:asm}
\end{equation}
with
\begin{equation}
H(k_x,k_y)
=
\exp
\left[
iz
\sqrt{
k_0^2-k_x^2-k_y^2
}
\right],
\label{eq:asm_transfer}
\end{equation}
where $k_0=2\pi/\lambda$. The angular-spectrum formulation \cite{khare2015angular} is used as the quantitative forward model throughout the reconstruction, retaining the full wave-optical propagation of both channels.


\begin{figure}[ht]
\centering
\includegraphics[width=\linewidth]{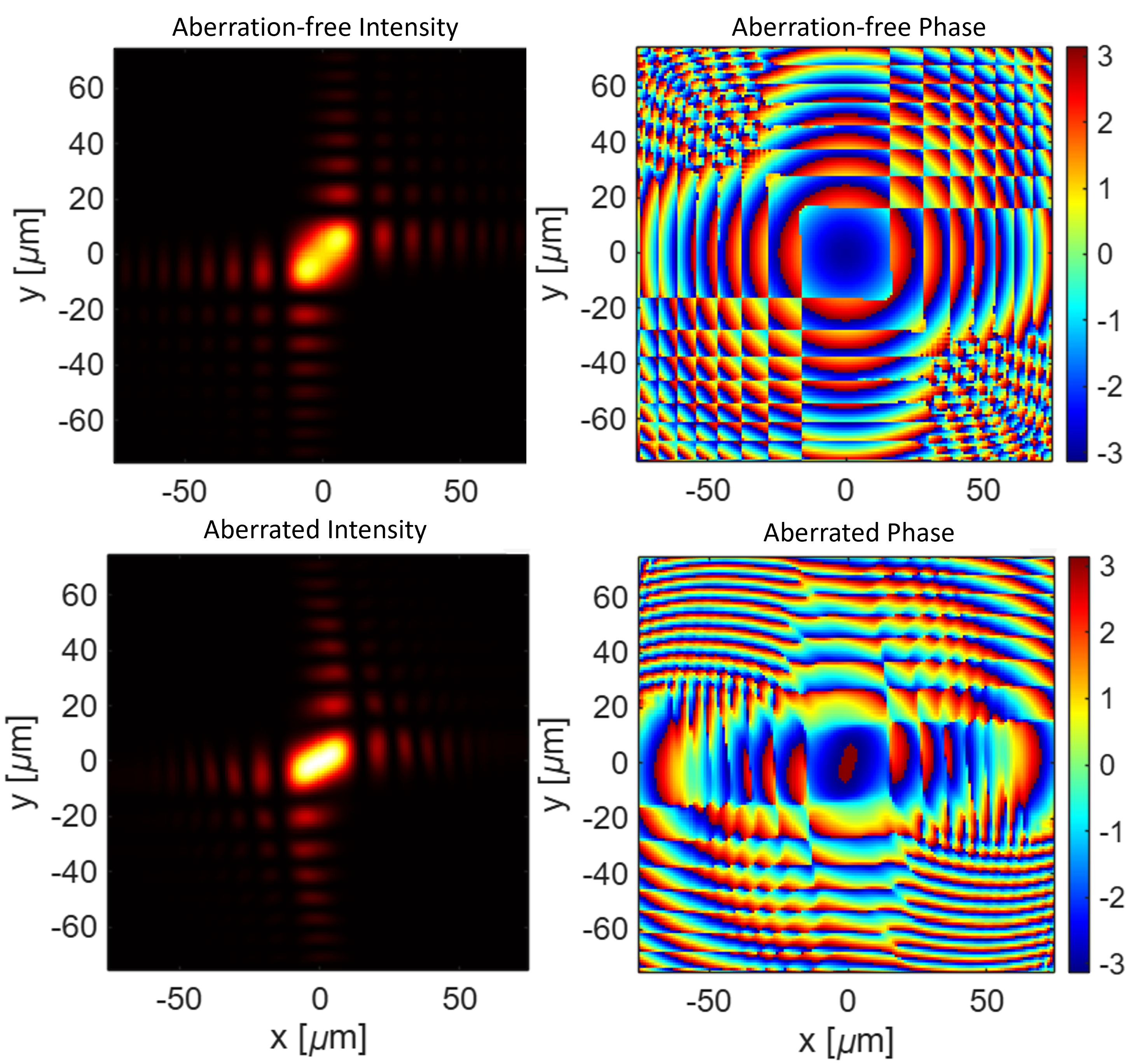}
\caption{
Interference of the conjugate cubic-phase channels without and with exit-pupil aberrations. The aberrated case includes $-2$ waves of defocus, $3$ waves of astigmatism, and $10$ waves of spherical aberration. A visible phase-shearing pattern appears in the interference phase in the presence of aberrations.
}
\label{fig:interference}
\end{figure}

\textit{Stationary-phase mapping:} The physical interpretation of the sensing mechanism follows from the stationary-phase condition. For one-dimensional notation, the propagated field associated with a cubic phase can be written in the form
\begin{equation}
U_{\pm}(x,z)
\propto
\int
E_{\mathrm{p}}(x')
\exp
\left[
i\Psi_{\pm}(x';x,z)\right]
dx',
\label{eq:stationary_integral}
\end{equation}
where, under the paraxial approximation,
\begin{equation}
\Psi_{\pm}
=
\Phi_{\mathrm{p}}(x')
\pm
\alpha\left(\frac{x'}{R}\right)^3
-
\frac{k_0(x-x')^2}{2z}.
\label{eq:stationary_phase}
\end{equation}
The dominant contribution to the propagated field is associated with locations satisfying
\begin{equation}
\frac{\partial\Psi_{\pm}}{\partial x'}=0.
\label{eq:stationary_condition}
\end{equation}
The cubic term introduces an opposite spatial mapping for the two channels. In the corresponding two-dimensional formulation, the pupil coordinates are mapped to different locations in the Fourier plane according to the sign of the cubic phase. Figure~\ref{fig:interference} provides a direct numerical visualization of this effect: in the presence of the specified pupil aberrations, a phase-shearing pattern is visible in the interference phase.

Denoting the corresponding mapped coordinates by $\mathbf{r}_{+}$ and $\mathbf{r}_{-}$, the relative phase can be interpreted locally as
\begin{equation}
\Delta\psi(\mathbf{r})
\approx
\Phi_{\mathrm{p}}(\mathbf{r}_{+})
-
\Phi_{\mathrm{p}}(\mathbf{r}_{-})
+
\Delta\psi_{0}(\mathbf{r}),
\label{eq:phase_shear}
\end{equation}
where $\Delta\psi_{0}$ represents the calibrated phase difference in the absence of aberration.

For a sufficiently small effective shear
\begin{equation}
\mathbf{s}
=
\mathbf{r}_{+}-\mathbf{r}_{-},
\end{equation}
Eq.~\eqref{eq:phase_shear} reduces locally to
\begin{equation}
\boxed{
\Delta\psi-\Delta\psi_{0}
\approx
\mathbf{s}\cdot
\nabla\Phi_{\mathrm{ab}}
}
\label{eq:gradient_shear}
\end{equation}
which provides a phase-shearing interpretation of the measurement. Equation~\eqref{eq:gradient_shear} is used as a physical interpretation of the sensing mechanism rather than as the primary reconstruction model. The exact numerical reconstruction instead retains the complete propagation of the common aberrated pupil in Eq.~\eqref{eq:asm}, thereby accounting for finite apertures, Gaussian apodization, diffraction, and the full spatial structure of the Airy fields.


\begin{figure}[ht]
\centering
\includegraphics[width=\linewidth]{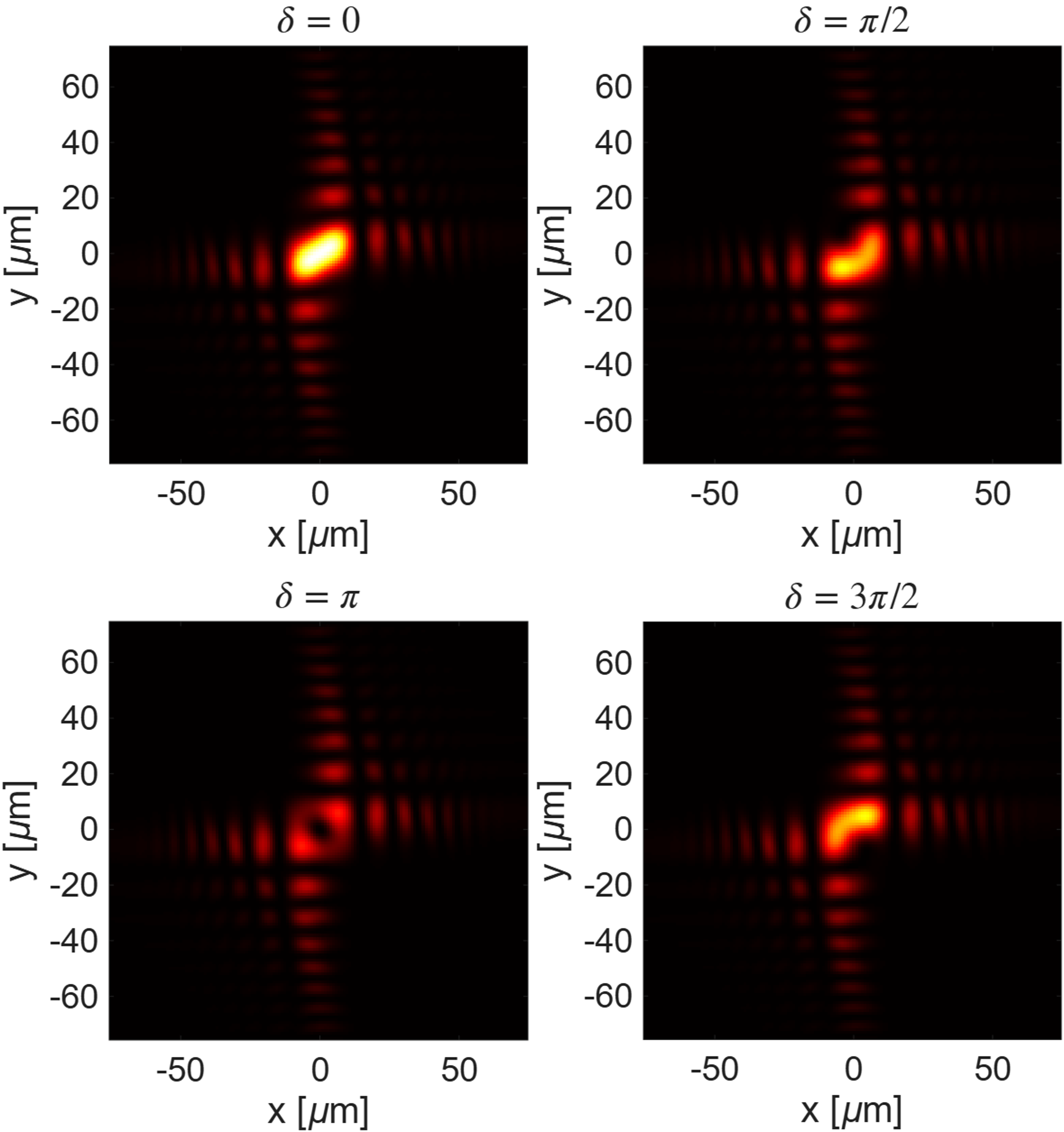}
\caption{
Four-step interferometric phase retrieval in the Fourier plane for a wavefront containing $10\lambda$ spherical aberration, $-2\lambda$ defocus, and $3\lambda$ astigmatism. Four intensity measurements are acquired with controlled relative phase shifts of $0$, $\pi/2$, $\pi$, and $3\pi/2$. These four phase-shifted measurements form the standard four-step phase-shifting interferometry scheme for recovering the relative phase through quadrature combinations \cite{schwider1993new}. The resulting quadratures provide the wrapped relative phase and the interference modulation used to define a valid reconstruction region.
}
\label{fig:phase_retrieval}
\end{figure}

\textit{Common-path interference and four-step phase retrieval:} The two propagated fields are coherently combined in the Fourier plane. Writing
\begin{equation}
U_{+}(\mathbf{r})
=
A_{+}(\mathbf{r})
e^{i\psi_{+}(\mathbf{r})},
\qquad
U_{-}(\mathbf{r})
=
A_{-}(\mathbf{r})
e^{i\psi_{-}(\mathbf{r})},
\label{eq:polar_fields}
\end{equation}
the detected field for a controlled phase shift $\delta$ is
\begin{equation}
U_{\delta}(\mathbf{r})
=
U_{+}(\mathbf{r})
+
e^{i\delta}
U_{-}(\mathbf{r}).
\label{eq:combined_field}
\end{equation}
The corresponding intensity is
\begin{equation}
I_{\delta}(\mathbf{r})
=
A_{+}^{2}
+
A_{-}^{2}
+
2A_{+}A_{-}
\cos
\left[
\Delta\psi(\mathbf{r})-\delta
\right],
\label{eq:interference_intensity}
\end{equation}
where
\begin{equation}
\Delta\psi(\mathbf{r})
=
\psi_{+}(\mathbf{r})
-
\psi_{-}(\mathbf{r}).
\label{eq:relative_phase}
\end{equation}

Four interferograms are acquired using
\begin{equation}
\delta
=
0,\quad
\frac{\pi}{2},\quad
\pi,\quad
\frac{3\pi}{2}.
\end{equation}
This four-step phase-shifting scheme provides the standard quadrature measurements required for direct recovery of the wrapped relative phase \cite{schwider1993new}. The phase quadratures are then
\begin{equation}
Q_c
=
I_{0}-I_{\pi},
\label{eq:quadrature_cos}
\end{equation}
and
\begin{equation}
Q_s
=
I_{\pi/2}-I_{3\pi/2}.
\label{eq:quadrature_sin}
\end{equation}
Substitution of Eq.~\eqref{eq:interference_intensity} gives
\begin{equation}
Q_c
=
4A_{+}A_{-}\cos\Delta\psi,
\end{equation}
and
\begin{equation}
Q_s
=
4A_{+}A_{-}\sin\Delta\psi.
\end{equation}
The wrapped relative phase is consequently obtained without requiring knowledge of the individual field amplitudes:
\begin{equation}
\boxed{
\Delta\psi_{\mathrm{wrap}}
=
\operatorname{atan2}
\left(
I_{\pi/2}-I_{3\pi/2},
I_{0}-I_{\pi}
\right)
}
\label{eq:four_step_phase}
\end{equation}
The corresponding interference modulation is
\begin{equation}
M(\mathbf{r})
=
2A_{+}(\mathbf{r})A_{-}(\mathbf{r}),
\label{eq:modulation}
\end{equation}
which provides a natural measure of the local interferometric signal strength. Pixels with insufficient modulation are excluded from subsequent phase analysis.


\begin{figure}[ht]
\centering
\includegraphics[width=\linewidth]{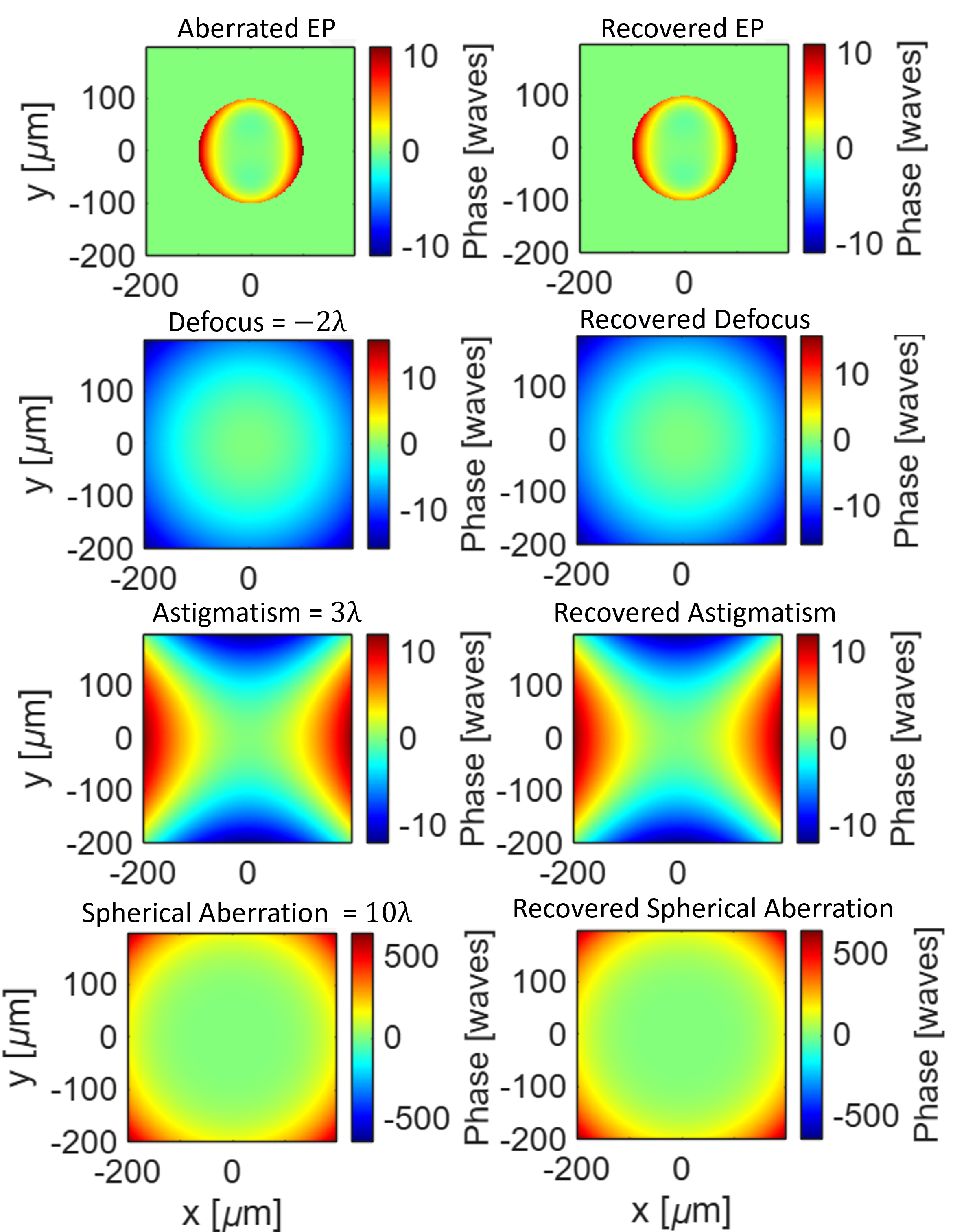}
\caption{
Inverse aberration recovery. The measured four-step interferometric data are used in an inverse forward-model optimization to recover the exit-pupil aberration coefficients. Defocus, astigmatism, and spherical aberration are recovered as $-2.00000$, $3.00000$, and $9.99998$ waves, respectively, compared with the true values of $-2$, $3$, and $10$ waves. The resulting exit-pupil phase is reconstructed with an RMS phase error of $<10^{-6}$ rad, demonstrating accurate recovery of the imposed aberration.
}
\label{fig:reconstruction}
\end{figure}

\textit{Inverse aberration recovery:} The four-step procedure provides the measured relative phase in the Fourier plane, but the measured phase is not assumed to be identical to the pupil aberration. Because propagation occurs after the common aberrated pupil and cubic-phase encoding, the relationship between the unknown pupil phase and the measured phase is generally nonlocal. Accordingly, aberration recovery is formulated as an inverse problem using the complete wave-optical forward model.

The unknown aberration is represented by a parameter vector
\begin{equation}
\mathbf{p}
=
[W_{\mathrm{DEF}},W_{\mathrm{AST}},W_{\mathrm{SA}},\ldots]^{T},
\label{eq:parameter_vector}
\end{equation}
where the parameters represent selected low-order aberration coefficients. For example, the aberration phase may be expressed as
\begin{equation}
\Phi_{\mathrm{ab}}(\rho,\theta)
=
2\pi
\left[
W_{\mathrm{DEF}}\rho^2
+
W_{\mathrm{AST}}\rho^2
\cos 2(\theta-\theta_{\mathrm{AST}})
+
W_{\mathrm{SA}}\rho^4
\right],
\label{eq:aberration_model}
\end{equation}
where $\rho=r/R$ is the normalized pupil radius.

For a trial parameter vector $\mathbf{p}$, the model generates the common aberrated pupil field,
\begin{equation}
E_{\mathrm{p}}(\mathbf{r};\mathbf{p})
=
A(\mathbf{r})E_{\mathrm{G}}(\mathbf{r})
\exp
\left[
i\Phi_{\mathrm{p}}(\mathbf{r};\mathbf{p})\right],
\label{eq:model_pupil}
\end{equation}
followed by the two cubic-phase channels
\begin{equation}
U_{\pm}(\mathbf{r};\mathbf{p})
=
\mathcal{P}_{z}
\left\{
E_{\mathrm{p}}(\mathbf{r};\mathbf{p})
 e^{\pm i\Phi_c(\mathbf{r})}
\right\}.
\label{eq:model_channels}
\end{equation}
The four predicted interferograms are then
\begin{equation}
I_{\delta}^{\mathrm{model}}
(\mathbf{r};\mathbf{p})
=
\left|
U_{+}(\mathbf{r};\mathbf{p})
+
e^{i\delta}
U_{-}(\mathbf{r};\mathbf{p})
\right|^{2}.
\label{eq:model_interferograms}
\end{equation}
The aberration parameters are recovered by minimizing the discrepancy between the measured and modeled interferograms,
\begin{equation}
\mathcal{L}(\mathbf{p})
=
\sqrt{
\frac{1}{4N_v}
\sum_{\delta}
\sum_{\mathbf{r}\in\Omega}
\left[
I_{\delta}^{\mathrm{model}}(\mathbf{r};\mathbf{p})
-
I_{\delta}^{\mathrm{meas}}(\mathbf{r})
\right]^2
},
\label{eq:loss}
\end{equation}
where $\Omega$ denotes the valid Fourier-plane region and $N_v$ is the number of valid pixels. This formulation avoids imposing a direct point-to-point identification between the measured relative phase and the pupil aberration. Instead, the complete diffraction process is incorporated into the inverse model.

For visualization, the recovered wrapped phase may be unwrapped using the modulation map in Eq.~\eqref{eq:modulation} as a quality measure. The primary quantitative result, however, is the recovery of the aberration parameters through the forward model.


\begin{figure}[ht]
\centering
\includegraphics[width=\linewidth]{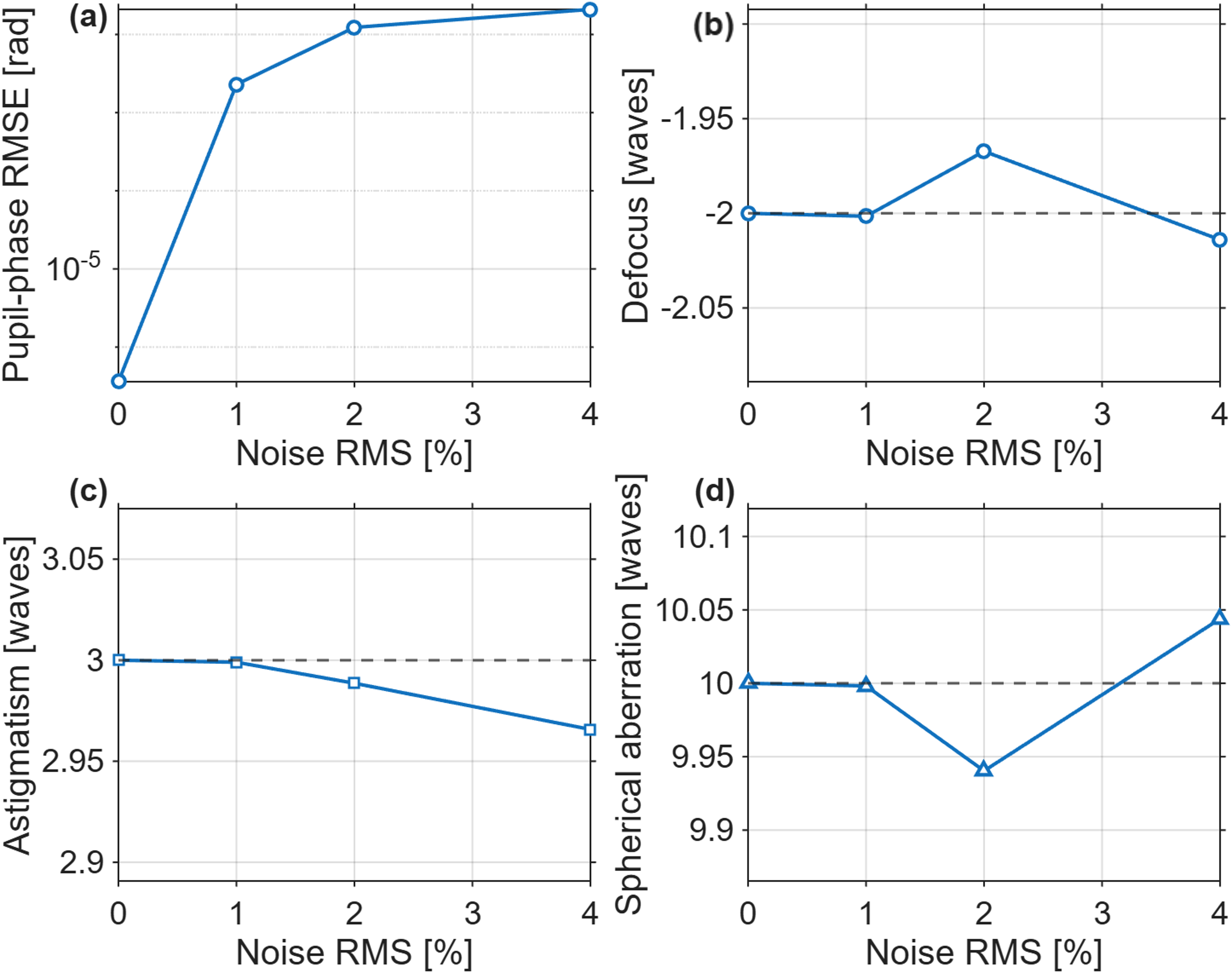}
\caption{
Noise tolerance of the proposed wavefront sensor. Reconstruction error is evaluated as a function of additive measurement noise applied independently to the four phase-shifted interferograms.
}
\label{fig:noise}
\end{figure}

\textit{Numerical validation and noise tolerance:} The proposed sensing framework is evaluated using wave-optical simulations of a Gaussian beam illuminating an aberrated exit pupil followed by conjugate cubic-phase encoding and angular-spectrum propagation. The numerical model uses
\begin{equation}
\lambda=633~\mathrm{nm},
\qquad
N=1024,
\qquad
L=1~\mathrm{mm},
\end{equation}
with a pupil radius of
\begin{equation}
R=100~\mu\mathrm{m},
\end{equation}
and a Gaussian beam waist of
\begin{equation}
w_0=50~\mu\mathrm{m}.
\end{equation}
The cubic-phase strength is selected such that the two channels form clearly distinguishable counter-accelerating Airy fields while maintaining substantial overlap in the Fourier-plane measurement region.

The reconstruction is first evaluated for individual aberration modes, including defocus, astigmatism, and spherical aberration, followed by a combined-aberration case. The recovered coefficients are compared with the known coefficients used to generate the simulated pupil. Figure~\ref{fig:reconstruction} summarizes the combined-aberration recovery. For the combined case, the imposed aberration consists of $-2$ waves of defocus, $3$ waves of astigmatism, and $10$ waves of spherical aberration, which are recovered with errors of less than $2\times10^{-5}$, $1\times10^{-5}$, and $2\times10^{-5}$ waves, respectively, in the noiseless simulation.

Wavefront reconstruction accuracy is quantified using the root-mean-square phase error,
\begin{equation}
\mathrm{RMSE}
=
\sqrt{
\frac{1}{N_p}
\sum_{\mathbf{r}\in P}
\left[
\Phi_{\mathrm{ab}}^{\mathrm{rec}}(\mathbf{r})
-
\Phi_{\mathrm{ab}}^{\mathrm{true}}(\mathbf{r})
\right]^2
},
\label{eq:rmse}
\end{equation}
where $P$ denotes the pupil region. To evaluate robustness, independent additive Gaussian noise is applied to each of the four measured interferograms before phase retrieval. The reconstruction is repeated for a range of noise levels, and the resulting aberration error is evaluated using Eq.~\eqref{eq:rmse}. Detailed numerical implementation and noise-model definitions are provided in Supplement 1.

Noise tolerance is evaluated by independently adding zero-mean Gaussian noise to each of the four phase-shifted interferograms before phase retrieval. Figure~\ref{fig:noise} shows the resulting pupil-phase RMSE and recovered aberration coefficients for noise RMS levels of $0$, $1$, $2$, and $4\%$ relative to the normalized interferogram intensity. The recovered coefficients remain close to their imposed values over this range, while the pupil-phase RMSE increases with noise level, demonstrating the expected degradation of reconstruction accuracy as measurement noise increases. In particular, at $4\%$ noise RMS, the recovered defocus, astigmatism, and spherical-aberration coefficients remain within approximately $0.02$, $0.04$, and $0.05$ waves, respectively, of their imposed values for the simulated realization. These results provide a numerical demonstration of noise tolerance for the proposed four-step reconstruction.


\textit{Discussion:} The proposed framework combines two complementary descriptions of the measurement. The stationary-phase formulation provides an intuitive interpretation in which the opposite cubic phases generate complementary spatial mappings and encode the wavefront through a phase-shearing relationship. The full angular-spectrum model provides the quantitative forward operator required for aberration reconstruction.

This distinction is important because the propagation of the common aberrated field and the cubic phase encoding are generally nonseparable. In particular,
\begin{equation}
\mathcal{P}_{z}
\left\{
E_{\mathrm{p}} e^{\pm i\Phi_c}
\right\}
\neq
\mathcal{P}_{z}\left\{E_{\mathrm{p}}\right\}
\mathcal{P}_{z}\left\{e^{\pm i\Phi_c}\right\}.
\end{equation}
Consequently, the measured relative phase cannot in general be interpreted as a direct copy of the pupil aberration.

The common-path architecture also means that the two interfering fields originate from the same incident wavefront and share the same optical path before their conjugate cubic phase responses are generated. This provides a natural route toward compact interferometric sensing while avoiding a conventional two-arm interferometer.

The four-step retrieval separates the phase measurement from the subsequent inverse reconstruction. The measured intensities provide the local relative phase and modulation, while the forward model establishes the relationship between the measured Fourier-plane signal and the unknown pupil aberration. Figure~\ref{fig:phase_retrieval} illustrates the four-step measurement, and Fig.~\ref{fig:reconstruction} shows the subsequent model-based recovery.

The present results demonstrate the feasibility of the proposed approach for the aberration conditions considered here. A natural direction for future work is single-shot phase recovery, which could further simplify the measurement by eliminating the need for sequential phase-shifted acquisitions. Single-frame computational recovery of complex optical fields has previously been demonstrated in digital holography \cite{khare2013single}, suggesting a possible route toward replacing the present sequential phase-shifting measurement with a single recorded interferogram. Extending this concept to the present counter-accelerating Airy-beam architecture will require a formulation that encodes the required phase information within a single measurement. Further investigation of larger and more complex aberrations will also be important for characterizing the operating range and robustness of the reconstruction. These extensions will build on the proof-of-concept demonstration presented here and provide a more comprehensive assessment of the method.


\textit{Conclusion:} A theoretical framework for Airy-beam wavefront sensing has been developed in which an aberrated Gaussian-illuminated exit pupil is encoded with oppositely signed cubic phases to generate two coherent Airy channels. A stationary-phase analysis provides a phase-shearing interpretation of the resulting spatial mapping, while common-path interference in the Fourier plane enables four-step phase retrieval. The recovered interferometric phase is subsequently related to the unknown exit-pupil aberration through a full angular-spectrum forward model and inverse parameter estimation. Numerical validation of aberration recovery and measurement-noise tolerance demonstrates the feasibility of the proposed architecture as a compact interferometric approach to wavefront sensing under the conditions considered here. The formulation is based on coherent wave propagation and engineered cubic phase and therefore provides a theoretical foundation for extending the sensing concept beyond conventional optical wavefronts.

\begin{backmatter}

\bmsection{Funding} 
No funding was received for this work.

\bmsection{Acknowledgment}
The author acknowledges the support and resources provided by the Rochester Institute of Technology.

\bmsection{Disclosures}
The author declares no conflicts of interest.

\bmsection{Data availability}
Data underlying the results presented in this paper are not publicly available at this time but may be obtained from the author upon reasonable request.

\bmsection{Supplemental document}
See Supplement 1 for detailed derivations of the cubic-phase Airy fields, stationary-phase mapping, angular-spectrum propagation, four-step phase retrieval, aberration parameterization, and numerical implementation.

\end{backmatter}



\clearpage
\section*{Supplement 1}
\addcontentsline{toc}{section}{Supplement 1}

\subsection*{S1. Cubic-Phase Airy-Field Formation}

The Airy-beam structure follows from the cubic phase. For a one-dimensional field, consider the Fourier representation
\begin{equation}
U_{\pm}(x,0)
\propto
\int
\widetilde{E}(k_x)
\exp
\left[
\pm i\alpha k_x^3
+
ik_xx
\right]
dk_x.
\label{eq:supp_airy}
\end{equation}
For a slowly varying spectral envelope, the cubic phase produces the Airy functional form. The finite Gaussian illumination provides the spectral apodization required for a finite-energy realization \cite{siviloglou2007accelerating}.

For the two-dimensional cubic phase used in the main text,
\begin{equation}
\Phi_{c,\pm}(x,y)
=
\pm
\alpha
\left[
\left(\frac{x}{R}\right)^3
+
\left(\frac{y}{R}\right)^3
\right],
\end{equation}
the two transverse dimensions acquire opposite cubic phase contributions. The corresponding propagated fields therefore exhibit complementary Airy-type spatial mappings.

\subsection*{S2. Stationary-Phase Mapping}

For a propagated field written in the pupil coordinate $x'$, the paraxial phase is
\begin{equation}
\Psi_{\pm}(x')
=
\Phi_{\mathrm{p}}(x')
\pm
\alpha\left(\frac{x'}{R}\right)^3
-
\frac{k_0(x-x')^2}{2z}.
\end{equation}
The dominant contribution is obtained from
\begin{equation}
\frac{\partial\Psi_{\pm}}
{\partial x'}
=0.
\end{equation}
Thus,
\begin{equation}
\frac{\partial\Phi_{\mathrm{p}}}{\partial x'}
\pm
\frac{3\alpha x'^2}{R^3}
+
\frac{k_0(x-x')}{z}
=0.
\label{eq:supp_stationary}
\end{equation}
In the absence of aberration, the opposite signs of the cubic term generate complementary stationary-phase mappings. When an aberration is present, the additional phase gradient perturbs the mapping and consequently modifies the Fourier-plane phase structure.

For sufficiently small mapping separation, the two channels sample nearby regions of the aberrated wavefront and the relative phase can be approximated by
\begin{equation}
\Delta\psi
-
\Delta\psi_0
\approx
\mathbf{s}\cdot
\nabla
\Phi_{\mathrm{ab}}.
\end{equation}
This approximation provides the physical interpretation of the sensing mechanism, whereas the full numerical reconstruction uses the angular-spectrum propagation model.

\subsection*{S3. Angular-Spectrum Propagation}

The numerical propagation used in the simulations is
\begin{equation}
U_{\pm}(x,y,z)
=
\mathcal{F}^{-1}
\left\{
\mathcal{F}
\left[
E_{\mathrm{p}}(x,y)
 e^{\pm i\Phi_c(x,y)}
\right]
H(k_x,k_y)
\right\},
\end{equation}
where
\begin{equation}
H(k_x,k_y)
=
\exp
\left[
iz
\sqrt{k_0^2-k_x^2-k_y^2}
\right].
\end{equation}
Only propagating spatial-frequency components are retained in the numerical implementation. The spatial-frequency components satisfying $k_x^2+k_y^2>k_0^2$ are excluded from the propagated field.

\subsection*{S4. Four-Step Phase Retrieval}

The four measured intensities are
\begin{equation}
I_0
=
A_+^2+A_-^2
+
2A_+A_-\cos\Delta\psi,
\end{equation}
\begin{equation}
I_{\pi/2}
=
A_+^2+A_-^2
+
2A_+A_-\sin\Delta\psi,
\end{equation}
\begin{equation}
I_{\pi}
=
A_+^2+A_-^2
-
2A_+A_-\cos\Delta\psi,
\end{equation}
and
\begin{equation}
I_{3\pi/2}
=
A_+^2+A_-^2
-
2A_+A_-\sin\Delta\psi.
\end{equation}
Consequently,
\begin{equation}
\Delta\psi
=
\operatorname{atan2}
\left[
I_{\pi/2}-I_{3\pi/2},
I_0-I_{\pi}
\right].
\end{equation}
The corresponding modulation is
\begin{equation}
M
=
\frac{1}{2}
\sqrt{
\left(I_0-I_{\pi}\right)^2
+
\left(I_{\pi/2}-I_{3\pi/2}\right)^2
}.
\end{equation}
A modulation threshold is applied to define the valid reconstruction region.

\subsection*{S5. Aberration Parameterization}

The simulated aberration can be represented using a low-order basis,
\begin{equation}
\Phi_{\mathrm{ab}}
=
2\pi
\sum_j
p_j Z_j(\rho,\theta),
\end{equation}
where $Z_j$ denotes the selected normalized aberration basis function and $p_j$ is its coefficient.

For the numerical demonstrations, the parameter vector contains
\begin{equation}
\mathbf{p}
=
[W_{\mathrm{DEF}},W_{\mathrm{AST}},W_{\mathrm{SA}}]^T.
\end{equation}
The corresponding phase model is
\begin{equation}
\Phi_{\mathrm{ab}}(\rho,\theta)
=
2\pi
\left[
W_{\mathrm{DEF}}\rho^2
+
W_{\mathrm{AST}}\rho^2
\cos2(\theta-\theta_{\mathrm{AST}})
+
W_{\mathrm{SA}}\rho^4
\right].
\end{equation}
For the combined-aberration case shown in Figs.~\ref{fig:interference} and~\ref{fig:reconstruction}, the coefficients are $W_{\mathrm{DEF}}=-2$, $W_{\mathrm{AST}}=3$, $\theta_{\mathrm{AST}}=0$, and $W_{\mathrm{SA}}=10$, with coefficients expressed in waves. The forward model is evaluated for each trial parameter vector and compared with the measured four-step interferograms.

\subsection*{S6. Noise Model}

Additive Gaussian noise is independently applied to the four intensity measurements,
\begin{equation}
I_{\delta}^{\mathrm{noisy}}
=
I_{\delta}+n_{\delta},
\end{equation}
where
\begin{equation}
n_{\delta}
\sim
\mathcal{N}(0,\sigma_n^2).
\end{equation}
The noise level is normalized relative to the maximum measured intensity. Reconstruction performance is evaluated over a range of noise levels using the pupil-phase RMSE and recovered aberration coefficients.

\end{document}